# Competing spin-glass and spin-fluctuation states in $Nd_xPr_{4-x}Ni_3O_8$

Shangxiong Huangfu[1,2], Zurab Guguchia[3], Tian Shang[4], Hai Lin[2], Huanlong Liu[2], Xiaofu Zhang[5,6], Hubertus Luetkens[3], Andreas Schilling[2]

[1]Laboratory for High Performance Ceramics, Empa, Überlandstrasse 129, CH-8600 Dübendorf, Switzerland

[2]Department of Physics, University of Zurich, Winterthurerstrasse 190 CH-8057 Zurich Switzerland

[3]Laboratory for Muon Spin Spectroscopy (LMU), Paul Scherrer Institute (PSI), Forschungsstrasse 111, CH-5232 Villigen, Switzerland

[4]Key Laboratory of Polar Materials and Devices (MOE), School of Physics and Electronic Science, East China Normal University, Shanghai 200241, China

[5]State Key Laboratory of Functional Materials for Informatics, Shanghai Institute of Microsystem and Information Technology, Chinese Academy of Sciences (CAS), Shanghai 200050, China

[6]CAS Center for Excellence in Superconducting Electronics, Shanghai 200050, China

*Abstract*

Neodymium nickelates have attracted research interest due to their strongly correlated behaviour and remarkable magnetic properties. More importantly, superconductivity has recently been confirmed in thin-film samples of Sr-doped $NdNiO_2$, bringing the layered rare earth nickel oxides into the research spotlight. In this report, we present results on a series of $NdNiO_2$ analogues, $Nd_xPr_{4-x}Ni_3O_8$ ($x$ = 0.1, 0.25, 1, 2, and 4) obtained by topotactic reduction, in which we observe systematic changes in the magnetic behaviour. As the $Nd^{3+}$ content increases, the initially large spin-freezing region with magnetic frustration becomes smaller and gradually shifts to low temperatures, while the magnetic response gradually increases. The muon-spin spectroscopy measurements on $Nd_4Ni_3O_8$ show that this phenomenon is likely due to the enhancement of spin fluctuations in $Nd_xPr_{4-x}Ni_3O_8$, which weakens the spin frustration behaviour for high $Nd^{3+}$ contents and at low temperatures. These spin fluctuations can be caused by both Nd and Ni ions and could be one of the factors determining the occurrence of possible superconductivity.

## Introduction

Due to the strongly correlated $d$-electrons [1-3] and the anisotropic properties [4-7], quasi-two-dimensional (quasi-2D) nickelates had been considered to be good candidates for high-temperature superconductors (HTS) by Bednorz and Müller, even before they discovered the cuprate HTS [8]. With valence states between +2 and +1, the Ni ions in $T'$-type Nd nickelates ($Nd_{n+1}Ni_nO_{2n+2}$) show similar electron states of the outer shell $3d^{9-\delta}$ (mixed $3d^8$ and $3d^9$) to the hole-doped cuprate HTS [9-12]. These types of nickelates have quasi-2D crystal structures with Ni-O layers stacked by $n$ $NiO_2$ infinite planes, like the stacking of the $CuO_2$ layers in cuprate HTS [13,14]. Since $Nd^{3+}$ ions are relatively small compared to $La^{3+}$ and $Pr^{3+}$ ions, they can cause a larger distortion of the crystal structure and a change of the crystal field around the Ni ions than their La and Pr-containing homologues [15]. Moreover, compounds containing $Nd^{3+}$ ions can exhibit unique magnetic or/and strong electronic interactions [16-19]. $T'$-type Nd nickelates are therefore promising parent compounds for superconductors. Indeed, superconductivity has been observed in $Nd_{1-x}Sr_xNiO_2$ thin films, which have a hole concentration of 0.2 per $3d$-element, close to the situation in the hole-doped cuprate HTS [20,21,22].

The relatively low valence state ($\leq$ +2) of Ni and the large lattice distortion confer Nd nickelates of the $T'$-type pronounced metastability. Therefore, only a few corresponding Nd-containing compounds have been successfully synthesized. $NdNiO_2$ ($n = \infty$) has been extensively studied and shows a paramagnetic susceptibility that obeys a Curie-Weiss law above 55 K. Data from powder neutron diffraction experiments do not indicate long-range antiferromagnetic (AFM) ordering [23], while possible spin-wave excitations have been observed by resonant inelastic X-ray scattering measurements [24] and the behaviour of spin fluctuations have been revealed by Raman scattering studies [25]. Although superconductivity has been observed in thin films of the hole-doped compound $Nd_{1-x}Sr_xNiO_2$ [20,21], it is absent in bulk polycrystalline samples [26]. For bilayer ($n = 2$) compounds, only the La-substituted samples $Nd_{3-x}La_xNi_2O_6$ with a minimum $x$ of 1.75 have been described

[27]. The trilayer ($n = 3$) $Nd_4Ni_3O_8$ has an average Ni ion valence state of +1.33 which is quite close to that of the superconductor $Nd_{0.8}Sr_{0.2}NiO_2$ (+ 1.2) [28,29], while no superconductivity has been found even at high pressure [30,31]. Comprehensive optical measurements on $Nd_4Ni_3O_8$ show charge-stripe fluctuations [32]. A theoretical investigation predicts a Curie-Weiss-type magnetic behaviour of $Nd_4Ni_3O_8$ due to the predominant magnetic moments of the Nd ions, while the Ni-O bonds may, however, play an important role in other physical properties [33]. Moreover, Sm-substituted $Nd_{3.5}Sm_{0.5}Ni_3O_8$ samples show no change in their electrical behaviour as compared to unsubstituted samples, even at a high pressure up to 2 GPa [34]. Pr-substituted $Nd_xPr_{4-x}Ni_3O_8$ samples exhibit metallic behaviour and a positive correlation between resistivity and Nd content $x$ [35].

To gain insight into $T$-type Nd nickelates and to investigate the possibility of superconductivity in them, we have synthesized $Nd_4Ni_3O_8$ and their Pr-substituted isologues $Nd_xPr_{4-x}Ni_3O_8$ ($x = 0.1, 0.25, 1$ and $2$) by topotactic reduction from their $T$-type parent compounds [36]. Extensive magnetization measurements were carried out on these samples, including quasi-static (DC) and alternating current (AC) measurements. For $Nd_4Ni_3O_8$, we also performed heat capacity and a series of muon-spin rotation (μSR) experiments to further reveal its physical properties.

**Experimental**

First, we synthesized high-quality precursors of $Nd_xPr_{4-x}Ni_3O_{10}$ (x = 0.1, 0.25, 1, 2 and 4) by the method described in [37] (see Supplementary Material). The $Nd_xPr_{4-x}Ni_3O_8$ samples were then obtained via topotactic reduction by annealing the corresponding $Nd_xPr_{4-x}Ni_3O_{10}$ in 10% $H_2/N_2$ gas at 350 °C for 18 hours. All samples were characterized at room temperature by powder X-ray diffraction (XRD, Fig. 1a), and the results show sharp peaks with no visible impurities. The elemental composition of all samples was confirmed by energy dispersive spectroscopy (EDX, see Supplementary Material). Both DC and AC magnetizations were investigated using a Magnetic Properties Measurement System (MPMS 3, Quantum Design Inc.), which has an option for AC magnetic measurements (see Supplementary Material for details). The heat capacity measurements were carried out with the Physical Properties Measurement System (PPMS, Quantum Design Inc.) in the temperature range between 2 K and 300 K. Both the zero-field (ZF) and longitudinal-field (LF)μSR experiments were performed with the General Purpose Surface-Muon Instrument (GPS, [38]) at the Swiss Muon Source of the Paul Scherrer Institute, Switzerland.

**Results**

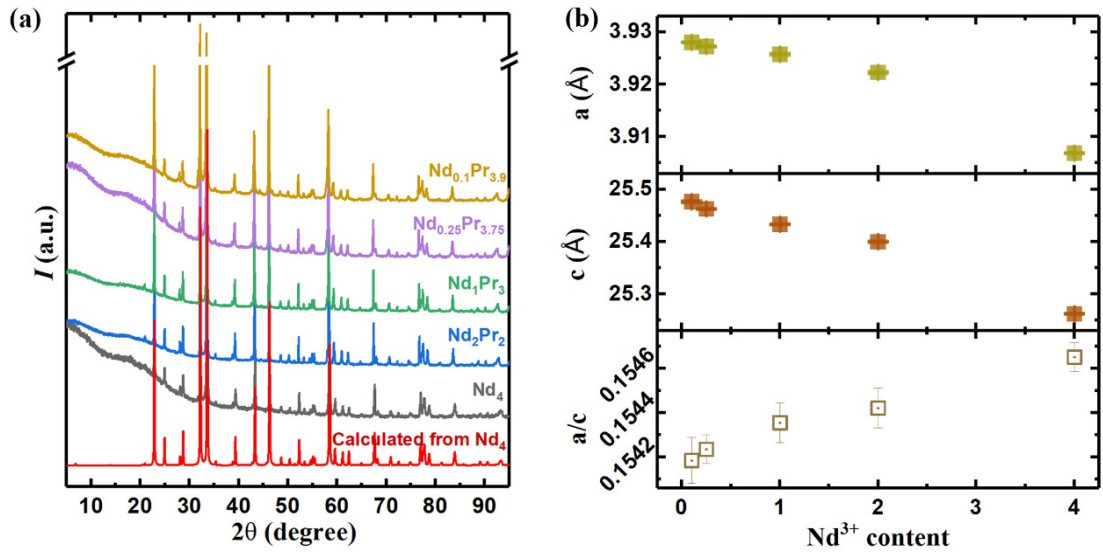

Fig. 1 (a) XRD patterns of $Nd_{4-x}Pr_xNi_3O_8$ samples; the red line corresponds to the calculated pattern for $Nd_4Ni_3O_8$; (b) Lattice parameters of $Nd_xPr_{4-x}Ni_3O_8$ as functions of the $Nd^{3+}$ content.

Compared to the original $T$-type $Ln_4Ni_3O_{10}$ structure (space group $P2_1/a$, monoclinic), the $T'$-type $Ln_4Ni_3O_8$ structure exhibits a higher symmetry with a tetragonal space group ($I4/mmm$). The holistic shift in the powder XRD patterns of $Nd_xPr_{4-x}Ni_3O_8$ reveals a decrease of the lattice parameters with increasing $x$, in agreement with Ref. 32, confirming the successful substitution. We refined these XRD data by a Rietveld analysis using the single crystal diffraction data of $Ln_4Ni_3O_8$ [12]. From the modeling results, all Ni and O ions within the infinite $NiO_2$ planes are in a regular square arrangement with no significant displacement, implying that there is no significant distortion within the NiO$_2$ planes. Despite the shrinkage of $a$ and $c$ with increasing Nd$^{3+}$ content (Fig. 1b), the $a/c$ ratio increases from ~ 0.1542 to ~ 0.1547 (~ 0.1541 for Pr$_4$Ni$_3$O$_8$ [36]).

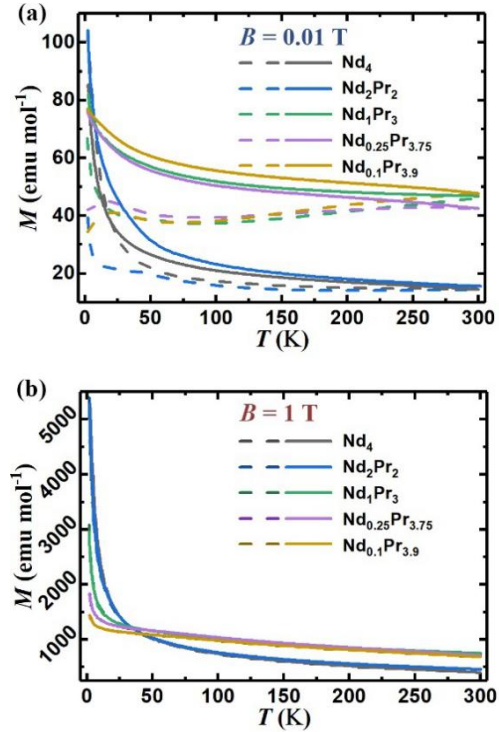

Fig.2 Temperature-dependent ZFC (dashed lines) and FC (solid lines) magnetizations of Nd$_{4-x}$Pr$_x$Ni$_3$O$_8$ in the external magnetic fields of 0.01 T (a) and 1 T (b), respectively.

Like Pr$_4$Ni$_3$O$_8$ [36], the magnetization data $M(T)$ of Nd$_x$Pr$_{4-x}$Ni$_3$O$_8$ samples exhibit different temperature dependences (see Fig. 2 and Supplementary Material) in different magnetic fields $B$. The $M(T)$ curves in fields below 0.5 T show a strong dependence on the magnetic history. The zero-field cooling (ZFC) and field cooling (FC) curves gradually separate in the low-temperature region. The irreversible region increases with decreasing field, and once 0.01 T is reached, the two curves are completely separated throughout the measured temperature range (2 K to 300 K). In contrast, the two curves do not differ between 2 K and 300 K above 0.5 T. Unlike Pr$_4$Ni$_3$O$_8$, the magnetization of Nd$_x$Pr$_{4-x}$Ni$_3$O$_8$ shows a significant sharp increase at low temperatures, especially at high magnetic fields. The magnetization in a field of $B$ = 1 T at 2 K increases from ~1.4 ×10$^3$ emu mol$^{-1}$ (for $x$ = 0.1) to ~5.3 ×10$^3$ emu mol$^{-1}$

(for $x$ = 4), compared to $Pr_4Ni_3O_8$ with ~1.2 ×10$^3$ emu mol$^{-1}$ [36], implying a stronger magnetic response for Nd-richer compounds.

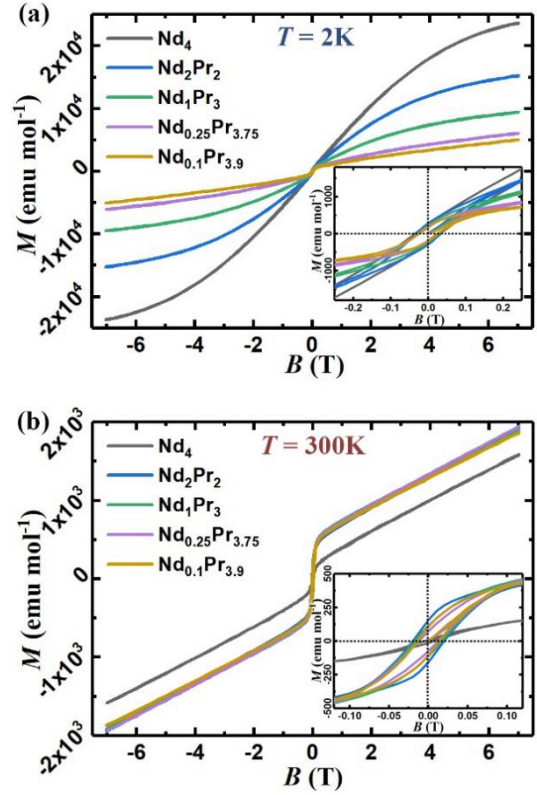

Fig.3 Isotherm field-dependent magnetizations $M(B)$ for magnetic fields $B$ bewteen -7 T and 7 T and temperatures $T$ at 2 K (a) and 300 K (b), respectively. The insets are enlarged views showing the hysteretic magnetic behaviour.

The magnetic field-dependent magnetizations $M(B)$ of $Nd_xPr_{4-x}Ni_3O_8$ at 2 K and 300 K are shown in Fig. 3a and b, respectively. The insets in Fig. 3 show typical hysteresis loops at both 2 K and 300 K, indicating ferromagnetic (FM)-like behaviour at all measured temperatures, similar to $Pr_4Ni_3O_8$ [36]. Instead of a pure saturation in the high-field limit, we observe an additional linear field dependence of the magnetization for all samples at high temperatures. However, at low temperatures, a clear linear field dependence of the magnetization could be separated from the saturated value in the high-field region only for the samples with $x$ = 0.1, 0.25, and 1, while for the samples with high $Nd^{3+}$ content ($x$ = 2 and 4) the magnetizations only tend to saturate without completion. As shown in Fig. 3a, the magnetizations show a significant positive correlation with the $Nd^3$ content $x$, from ~ 5.0 ×10$^3$ emu mol$^{-1}$ to ~ 2.37 ×10$^4$ emu

mol$^{-1}$ in 7 T at 2 K.

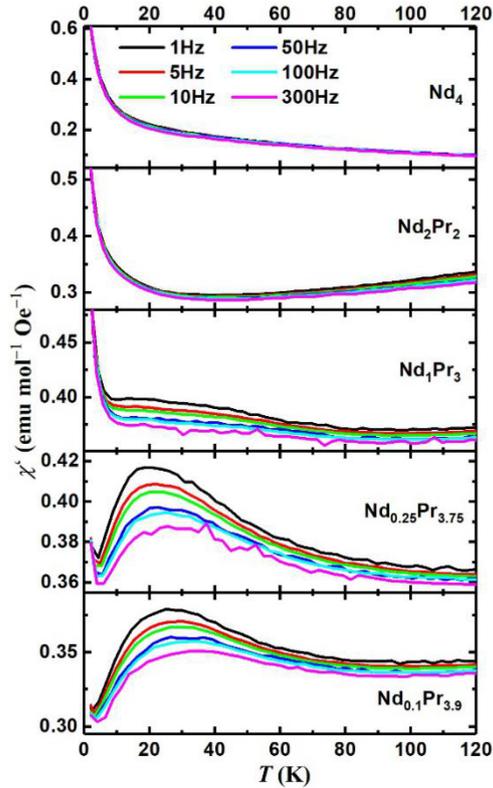

Fig.4 Real part ($\chi'$) of the AC magnetic susceptibility vs. temperature, measured with frequencies ranging from 2 Hz to 300 Hz, for Nd$_4$Ni$_3$O$_8$ (a), Nd$_2$Pr$_2$Ni$_3$O$_8$ (b), Nd$_1$Pr$_3$Ni$_3$O$_8$ (c), Nd$_{0.25}$Pr$_{3.75}$Ni$_3$O$_8$ (d), Nd$_{0.1}$Pr$_{3.9}$Ni$_3$O$_8$ (e), respectively.

Considering that Pr$_4$Ni$_3$O$_8$ has been reported to exhibit spin-glass behaviour [36], we have performed AC magnetization measurements on the Nd$_x$Pr$_{4-x}$Ni$_3$O$_8$ series in a zero static external magnetic field. Fig. 4 shows the temperature dependence of the real part $\chi'(T)$ of the AC susceptibility with AC frequencies from 1 Hz to 300 Hz in the temperature range between 2 K and 120 K. In general, all Nd$_x$Pr$_{4-x}$Ni$_3$O$_8$ compounds show a negative frequency dependence of $\chi'$, which can be considered as an indication of a non-equilibrium state with frustrated magnetization, e.g., a spin-glass state. The spin-glass freezing temperatures for Nd$_x$Pr$_{4-x}$Ni$_3$O$_8$ (x = 0.1, 0.25, and 1) can be determined from the respective maxima in $\chi'(T)$ [36,39,40]. These maxima shift toward higher temperatures with increasing AC frequency. With decreasing x, the "peak" temperatures (at 1 Hz) decrease from ≈ 27 K to ≈ 15 K and are well below those of $Pr_4Ni_3O_8$ (≈ 70 K) [36]. However, for $Nd_xPr_{4-x}Ni_3O_8$ samples with $x$ = 2 and 4, we observe hardly any peaks in the corresponding $\chi'(T)$ data. Moreover, the dependence of $\chi'$ on the AC frequency becomes considerably weaker with increasing $Nd^{3+}$ content. A clear spin-glass behaviour as in $Pr_4Ni_3O_8$ at low temperatures, i.e., the shift and decrease of $\chi'$ with AC frequency, is gradually replaced by a strong, frequency-independent increase of $\chi'$ as $Pr^{3+}$ is successively replaced by $Nd^{3+}$. In agreement with the $M(T)$ results, this low-temperature increase with $x$ can be taken as an indication for different magnetic state emerging below ≈ 3 K - 6 K for $Nd_xPr_{4-x}Ni_3O_8$ with $x$ = 1, 2, and 4, in which frustrated magnetic behaviour is absent and the magnetic susceptibility is enhanced.

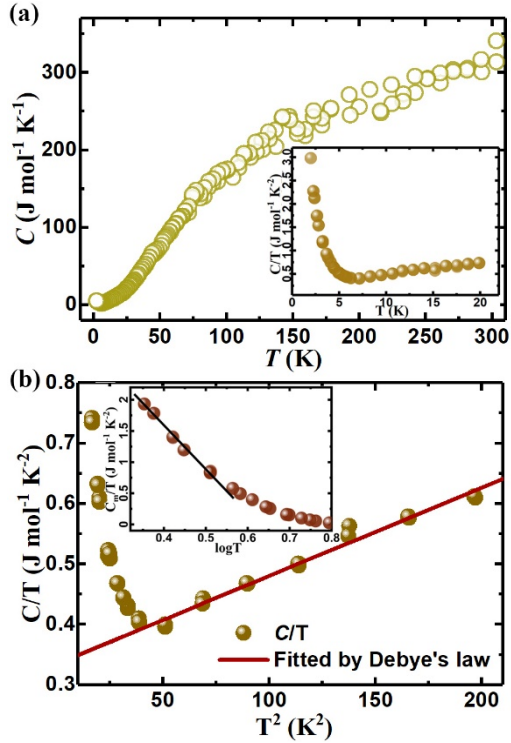

Fig.5 (a) Temperature dependence of the heat capacity of $Nd_4Ni_3O_8$ between 2 K and 300 K; the inset shows $C/T$ in the low temperature range; (b) $T^2$ dependence of $C/T$ from 4 K to 15 K; the red line corresponds to a fit to the Debye model; the inset shows the magnetic contribution $C_m/T$ vs. $\log T$ in the low temperature range.

To further investigate the physical properties of the end member ($x$ = 4) of the series, $Nd_4Ni_3O_8$, we measured the heat capacity at temperatures between 2 K and 300 K

(Fig. 5a). The zero-field heat capacity $C(T)$ shows no obvious peaks or kinks throughout the whole measurement range, reaching a value of ~ 340 Jmol$^{-1}$ K$^{-1}$ at 300 K, which is close to the classical Dulong-Petit value ($C_V = 3n$R ≈ 374 Jmol$^{-1}$ K$^{-1}$, where $R$ = 8.314 Jmol$^{-1}$ K$^{-1}$ and $n$ is the number of atoms per unit cell), and it is similar to the homologues Pr$_4$Ni$_3$O$_8$ and La$_4$Ni$_3$O$_8$ [12]. Below 6 K (inset in Fig. 5a), the heat capacity increases with decreasing temperature, which has also been observed in similar layered Nd-nickelates such as NdNiO$_3$ [41] and Nd$_4$Ni$_3$O$_{10}$ [19]. Above 6 K, as shown in Fig. 5b, the temperature-dependent $C/T$ can be well fitted by the Debye model for low temperatures, $C/T = \gamma + \beta T^2$, where $\gamma$ represents the electron and $\beta T^2$ the phonon contribution to the heat capacity, respectively. Fitting the experimental data gives $\beta \approx 1.46$ mJ mol$^{-1}$K$^{-4}$, corresponding to a Debye temperature of $\Theta_D \approx 270$ K, and $\gamma \approx 334$ mJ mol$^{-1}$K$^{-2}$. These values are comparable to those reported for the corresponding $T$-type compound Nd$_4$Ni$_3$O$_{10}$, with $\Theta_D \approx 300$ K and $\gamma \approx 146$ mJ mol$^{-1}$K$^{-2}$ [19]. After subtracting the electron and phonon contributions, we obtain the magnetic heat capacity ($C_m$) at low temperatures. The plot of $C_m/T$ vs. logT in the inset of Fig. 5b seems to indicate a linear dependence at low temperatures, which could point to a 3D ferromagnetic-interaction scenario in spin fluctuation theories, as it has been applied to many systems with strong correlations [42].

Positive muons can serve as an extremely sensitive local probe for detecting small internal magnetic fields, spin fluctuations and ordered magnetic volume fractions in the bulk materials. µSR is a particularly powerful tool for investigating materials with inhomogeneous magnetic states. We performed µSR measurements to investigate the complex magnetic nature of Nd$_4$Ni$_3$O$_8$ covering a wide temperature range. Fig. 6a shows the ZF-µSR spectra at representative temperatures. The absence of oscillations indicates the absence of static long-range magnetic order in Nd$_4$Ni$_3$O$_8$ down to 1.5 K. However, it shows that the muon spin relaxation has a clearly observable temperature dependence. Namely, there is an increase of the relaxation rate, which is similar to Pr$_4$Ni$_3$O$_8$ [36] but different from La$_4$Ni$_3$O$_8$ [43]. The fast depolarization of the µSR signal could be either due to a wide distribution of static fields, and/or to strongly fluctuating magnetic moments. To discriminate between these two possibilities, we carried out μSR experiments under a magnetic field applied longitudinally to the muon spin polarization. As can be seen in the longitudinal field (LF) -μSR spectra (Figs. 6b and c), polarization can be recovered (though not complete) by the application of a small external longitudinal magnetic field, $B$ = 5 mT, while at 5 K the typical depolarization of the μSR spectra can still be observed even in a relatively large field of $B$ = 300 mT, indicating the existence of strongly dynamic internal fields in the system.

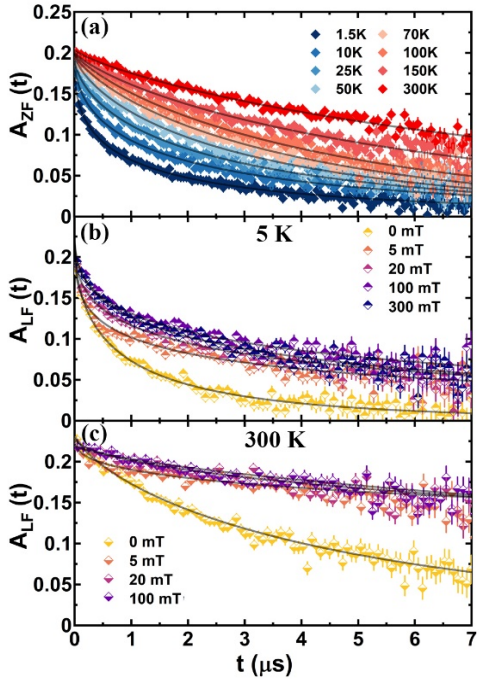

Fig.6 (a) the zero field (ZF)-$\mu$SR spectra from 1.5 K to 300 K; (b) and (c) the longitudinal field (LF)-$\mu$SR spectra at 5 K and 300 K, respectively.

The data in the entire temperature range are well modeled by a stretched exponential relaxation function, which is normally used to describe a system with a magnetic disorder (e.g., spin freezing) [44, 45, 46],

$$A(t) = A_0 + A_{relax} \exp(-\lambda t)^{\beta_\mu},$$

where $A_0$ and $A_{\text{relax}}$ are the non-relaxing and relaxing amplitudes, respectively; $\lambda$ represents the muon-spin relaxation rate, and $\beta_\mu$ is the so-called shape parameter indicating the shape of damping. This fitting assumes that the relaxation rate decreases with time as $t^{\beta_\mu - 1}$, which involves the sum of exponential and Gaussian terms and has a more physically tractable basis [44]. We will discuss the results of the analysis in the discussion section.

**Discussion**

To further analyze the DC magnetization in the $Nd_xPr_{4-x}Ni_3O_8$ series, we attempted to separate the paramagnetic contribution from the FM-like part, as is usually done for *T'*-type nickelates [36, 47, 48]. Considering that the FM-like contribution should be saturated at a sufficiently high magnetic field, we can extract an additional paramagnetic term that is proportional to the magnetic field *B* in the high-field region. From this, we can calculate the paramagnetic susceptibility $\chi_{para}$ (Fig. 7a) and thus obtain the saturated magnetic moment (Fig. 8) by subtracting the paramagnetic susceptibility from the total *M(B)* curve.

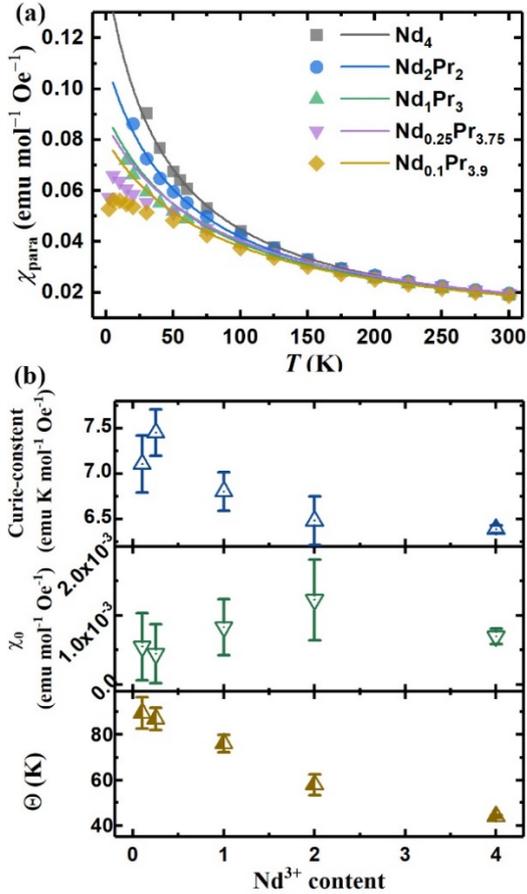

Fig.7 (a) Paramagnetic susceptibility of $Nd_xPr_{4-x}Ni_3O_8$ as calculated from the linear part of the $M(B)$ data and fitted by a Néel-type law; (b) The corresponding fitted values for various $Nd^{3+}$ contents.

For the compounds with low $Nd^{3+}$ contents (x < 2), paramagnetic susceptibilities can be determined by the above method because $M(B)$ shows a linear field dependence at high magnetic fields (> 4 T) throughout the whole temperature range from 2 K to 300 K (Fig. 3). However, for the compounds with $x = 2$ and 4, the low-temperature magnetization does not show a linear field dependence even in high magnetic fields, which is likely due to an incompletely saturated FM-like magnetization. Therefore, we cannot determine the paramagnetic susceptibilities of these samples at low temperatures. Similar to $Pr_4Ni_3O_8$, all these $\chi_{para}(T)$ curves in the high-temperature region can be well fitted by a Néel-type of the Curie-Weiss law, i.e., $\chi_{para}(T) = C/(T+\Theta)+\chi_0$ as shown in Fig. 7a, indicating a constant value of paramagnetic moments at high temperatures. The fitted Néel temperatures $\Theta$, Curie constants $C$, and Pauli paramagnetic susceptibilities $\chi_0$ are shown in Fig. 7b.

With increasing $Nd^{3+}$ content, the temperature range obeying the Néel-type law widens and shifts to lower temperatures from ≈ 100 K to ≈ 40 K, consistent with decreasing Néel temperatures $\Theta$ (from ≈ 90 K to ≈ 43 K), but much smaller than in the $Pr_4Ni_3O_8$ peer with the corresponding value of ≈ 160 K. This decrease in $\Theta$ also coincides with a slight decrease in the Curie constant $C$ and could therefore mark the onset of the spin-freezing process. Because of the theoretically similar magnetic moments of $Pr^{3+}$ and $Nd^{3+}$ (3.58 $\mu_B$ and 3.62 $\mu_B$, respectively), $C$ should be ≈ 6.4 emu K $Oe^{-1}$ $mol^{-1}$, provided that the paramagnetic susceptibilities stem solely from the rare earth ions. A comparison of the extracted $C$ values suggests that there must be an excess of magnetic moments in the Pr-rich samples. Since the inner shell electrons of $Pr^{3+}$ and $Nd^{3+}$ are insensitive to the environment, the variation of the Curie constant $C$ could be a consequence of a change in the magnetization of the Ni atoms caused by a change of the orbital electrons and the crystal field and ultimately to the structural changes shown in Fig. 1. The positive correlation between the lattice-parameter ratio

$a/c$ and the $Nd^{3+}$ content indicates an enhanced interaction between the $d_{3z^2-r^2}$ orbitals of Ni (which extend along the $c$ axis) with increasing $Nd^{3+}$ content, which in turn leads to an increase in the energy of these orbitals and eventually to an alteration of the crystal field around the Ni ions. All of this can cause a redistribution of electrons in the Ni $d$-orbitals, which may affect the magnetic moment and lead to an overall systematic change in $C$ and of other physical properties of $Nd_xPr_{4-x}Ni_3O_8$.

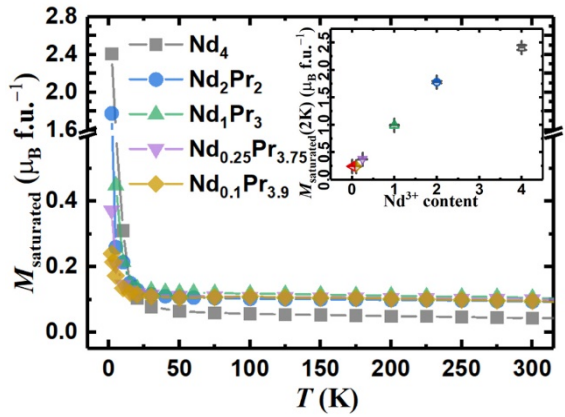

Fig.8 Saturated magnetic moment of the $Nd_{4-x}Pr_xNi_3O_8$ series obtained by subtracting the paramagnetic magnetizations from the $M(B)$ data. The inset shows the saturated magnetic moments at 2 K $vs.$ $Nd^{3+}$ content. The red triangle represents the value of $Pr_4Ni_3O_8$ for comparison.

We were able to reliably determine the magnetic moments of the saturation magnetizations ($M_{sat}$) for the compounds with $x < 2$ over the whole measured temperature range. For the compounds with $x = 2$ and 4, however, $M_{sat}$ could only be estimated from the $M(B)$ curves because the linear paramagnetic susceptibility could not be determined with sufficient precision at low temperatures. Since an intrinsic FM-like behaviour has already been demonstrated to be present in $Pr_4Ni_3O_8$ [36] and the powder XRD and µSR results did not reveal any impurities, we assume that the magnetic saturation at high temperatures is indeed intrinsic for $Nd_xPr_{4-x}Ni_3O_8$. As shown in Fig. 8, $M_{sat}$ of all $Nd_xPr_{4-x}Ni_3O_8$ compounds at high temperatures (i.e., above $\Theta$) is almost independent of temperature, while this quantity increases significantly in all compounds at low temperatures, being more pronounced for large $Nd^{3+}$ contents.

The $M_{sat}$ at 2 K varies from ≈ 0.25 μB/mol for $Nd_{0.1}Pr_{3.9}Ni_3O_8$ to ≈ 2.41 μB/mol for $Nd_4Ni_3O_8$, which is larger than in $Pr_4Ni_3O_8$ with a value of ≈ 0.24 μB/mol [36]. This enormous increase in $M_{sat}$ in the low-temperature range, indicating a strong magnetic interaction, is likely to have the same origin as the increases in the DC and AC magnetizations at low temperatures.

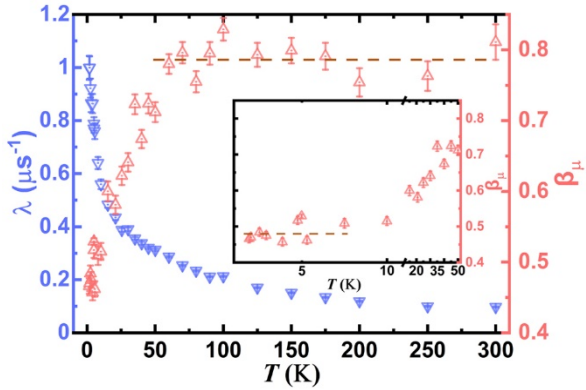

Fig.9 The relaxation rate λ and the shape parameter $β_μ$ of $Nd_4Ni_3O_8$ as obtained from a stretched exponential fit of the ZF-μSR spectra, as functions of temperature; the inset shows the low-temperature range of $β_μ$ while the red dashed line is to guide the eye.

At 300 K (Fig. 6c), the LF-μSR spectra of $Nd_4Ni_3O_8$ show a clear decoupling when a small external magnetic field (5 mT) is applied, pointing to a large fraction of static magnetization. As soon as the field increases to more than 20 mT, the μSR spectra is independent of the applied field, indicating a dynamic relaxation. Similarly to the 300 K data, the μSR spectra at 5 K (Fig. 6b) shows that the polarization is only partially decoupled by a magnetic field as high as 300 mT (the highest field applied), which suggests the presence of a dynamic magnetization, i.e., spin fluctuations, with disorder at low temperatures in $Nd_4Ni_3O_8$ [45,46,49,50], which is also in agreement with our heat capacity results and optical-spectroscopy data on $Nd_4Ni_3O_8$[31]. Unlike spin fluctuations associated with AFM order or spin-density waves, our data do not suggest a statically ordered magnetic pattern, and therefore again indicate a magnetic disorder. These spin fluctuations grow dramatically with decreasing temperature, leading to a strong magnetic response and an FM-like behaviour at low temperatures, which may overrun the frustration behaviour in the spin freezing process, thereby explaining the absence of frequency-dependent AC susceptibility in $Nd_4Ni_3O_8$ (Fig. 4).

From the fitted results of the stretched exponential function (Fig. 9), three different regimes can be clearly distinguished on the basis of the value of the shape parameter $\beta_\mu$: i) the temperature range from 60 K to 300 K with a nearly constant $\beta_\mu \approx 0.79$; ii) between 10 K and 50 K, where $\beta_\mu$ drops with decreasing temperature from $\approx 0.71$ to $\approx 0.52$; and iii) below 7.5 K, where $\beta_\mu$ again shows a nearly temperature-independent value of $\approx 0.49$ (inset of Fig. 9). Decrease of $\beta_\mu$ below 50 K indicates that the distribution of relaxation rates becomes broader below 50 K (so, becomes more disordered). In contrast to some spin-glass systems where $\beta_\mu$ is expected to decrease to about 1/3 [51], the corresponding values $\beta_\mu \sim 0.5$ of $Nd_4Ni_3O_8$ may point to somewhat different magnetic behaviour. The relaxation rate $\lambda$ shows a gradual increase from $\approx 0.1$ to $\approx 1.0$ $\mu s^{-1}$ throughout the whole temperature range from 2 K to 300 K upon cooling, and the increase becomes drastic once the lowest temperatures are reached (i.e., from $\approx 0.5$ to $\approx 1.0$ $\mu s^{-1}$ between 2 K and 10 K). The decrease of $\beta_\mu$ and the increase of $\lambda$ can always be observed during the spin-freezing process in a spin-glass system, where spin fluctuations slow down and frustration phenomena occur [52,53]. However, the continuing increase of $\lambda$ upon cooling further below 10 K, where the spin freezing process is already terminated, also indicates a different behaviour of $Nd_4Ni_3O_8$ than in normal spin-glass systems, where the relaxation rates always decrease below the respective freezing temperatures [41,54,55].

Comparing the DC magnetization data with the ZF- $\mu$SR results in $Nd_4Ni_3O_8$, we find that the Néel temperature ($\Theta \approx 43$ K) and the temperature of the steep drop in $\beta_\mu$ ($\approx 50$ K) roughly coincide, signaling the beginning of the spin-freezing process. In addition to the decrease of $\beta_\mu$, the spin freezing process between $\approx 10$ and $\approx 50$ K is confirmed by a slight frequency dependence of the magnetic AC susceptibility in this range (Fig. 4). Below 10 K, however, the FM-like behaviour increases, which can be attributed to the strengthening of the dynamic spin fluctuations. The magnetization becomes independent of frequency, although there may still be a magnetic disorder but with strong magnetic interactions.

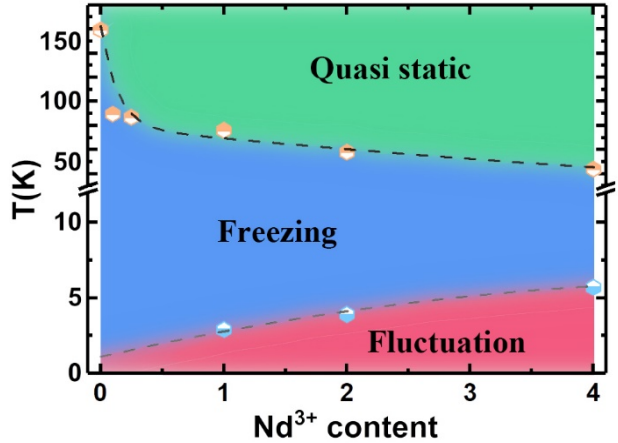

Fig.10 Schematic magnetic phase diagram of $Nd_xPr_{4-x}Ni_3O_8$ ($x$ = 0, 0.1, 0.25, 1, 2, and 4).

The gradual variation in terms of chemical composition, crystal structure, and magnetic behaviours in the $Nd_xPr_{4-x}Ni_3O_8$ ($x$ = 0, 0.1, 0.25, 1, 2, and 4) system suggests that the spin dynamics in this series also varies continuously as a function of $x$. Based on the results for $Nd_4Ni_3O_8$ ($x$ = 4) we can compare and distinguish the magnetic behaviours in the whole series, as we are summarizing in Fig. 10. At high temperatures, the ZF-μSR data for $x$ = 4 exhibit a temperature independent $\beta_\mu$ and small change in $\lambda$ (from ≈ 0.10 to ≈ 0.28 μs$^{-1}$), while the LF-μSR results show only a small amount of dynamic magnetization. Together with a nearly constant value of paramagnetic and FM-like configurations as deduced from the DC magnetization measurements for all samples, this may indicate a quasi-static magnetic state (green area in Fig. 10). We can define the low-temperature boundary of this regime by the fitted $\Theta$ values, which also mark the onset of the spin freezing process. Once the temperature reaches the value at which $\beta_\mu$ starts to drop in the $x$ = 4 sample, $\chi'$ becomes frequency dependent and a spin-freezing process sets in (entering the blue area in Fig. 10). As the temperature decreases further, the magnetic behaviour of all samples with high Nd content ($x = 1$, 2, and 4) shows the same dramatic change: the AC susceptibility becomes frequency independent, indicating the absence of frustrated magnetization, and also resulting in a constant $β_μ$ for the $x = 4$ sample. The temperatures, at which the AC susceptibility becomes frequency independent, therefore represent the upper boundary of this low-temperature area (red area in Fig. 10). The LF-μSR results of $Nd_4Ni_3O_8$ at 5 K suggest that the material is in a spin-fluctuation state with disorder. With increasing $Nd^{3+}$ content, the overall-temperature range showing spin freezing (blue area) gradually decreases.

We have previously reported that $Pr_4Ni_3O_8$ exhibits a complex spin-glass behaviour with typical frustrated magnetization, where the strong magnetic interaction over short distances ultimately leads to spin freezing [36]. The replacement of $Pr^{3+}$+ by $Nd^{3+}$ obviously alters the magnetic interaction, which may therefore lead to a transition from frustrated magnetization to dynamic magnetization at low temperatures, thereby changing the AC and DC magnetic behaviour. This spin fluctuations might be due to the interaction between Nd ions, which has been observed in previous reports [56, 57]. However, since a dynamic magnetization has also been observed in a non-rare-earth nickelate $LiNiO_2$ with similar $μ$SR spectra but with a triangular lattice symmetry ($Nd_4Ni_3O_8$ exhibits a tetragonal lattice symmetry) [58], we cannot rule out that the Ni-Ni interaction also plays an important role for the spin fluctuations. Due to a gradual shrinkage of the unit cell of $Nd_xPr_{4-x}Ni_3O_8$ and a positive trend of the $a/c$ value with increasing $x$, the interlayer spacing decreases with increasing $Nd^{3+}$ concentration, which may lead to an enhancement of the Ni-Ni interactions between the layers. This could also change the character of the spin fluctuations due to the Ni ions upon crossing over from a frustrated state to a strong dynamic state.

In summary, we conclude that the spin fluctuations in the Nd-rich samples and particularly in $Nd_4Ni_3O_8$ are of FM-like character. It has been reported that FM fluctuations can be observed in a variety of superconductors, including highly overdoped quasi-2D cuprate systems [59], heavy fermion systems [60,61], and some other superconductors (e.g., $Sr_2RuO_4$) [62]. Therefore, $Nd_4Ni_3O_8$ could be much closer to superconductivity than $Pr_4Ni_3O_8$. On the other hand, too strong magnetic interactions could counteract and suppress a superconducting state. Finally, it is worth noting that $LiNiO_2$, which exhibits a similar dynamic magnetization to $Nd_4Ni_3O_8$ at low temperatures, has been shown to be a candidate for a spin liquid, which justifies the interest in $Nd_4Ni_3O_8$ and in its ground state [58,63].

**Conclusions**

In this work, we have described the results of comprehensive magnetization, heat capacity, and a series of μSR measurements on the $Nd_xPr_{4-x}Ni_3O_8$ (x = 0.1, 0.25, 1, 2, and 4) system, which we synthesized by topotactic reduction from $Nd_xPr_{4-x}Ni_3O_{10}$. The magnetization measurements show that the spin-glass behaviour becomes gradually weaker with increasing $Nd^{3+}$ content and changes the magnetic behaviour at low temperatures. The μSR data from $Nd_4Ni_3O_8$ show the presence of spin fluctuations, that may surpass the spin freezing process and frustrated behaviour at low temperatures. These spin fluctuations could be attributed to both the magnetic moments of the Nd and Ni ions and hinder the formation of a potential superconducting ground state of these nickelates. Finally, we present a magnetic phase diagram of the $Nd_xPr_{4-x}Ni_3O_8$ series, which summarizes the magnetic behaviour of these compounds.